# Probing gas adsorption on individual facets of a metal nanoparticle

Pin Ann Lin[1,2], Jonathan Winterstein[1], John Kohoutek[1,2], Henri Lezec[1] and Renu Sharma[1*]

Affiliation:

[1]Center for Nanoscale Science and Technology, National Institute of Standards and Technology, Gaithersburg, MD 20899-6203, USA

[2] Maryland NanoCenter, University of Maryland, College Park, MD 20742, USA

*Corresponding author email:

Renu.sharma@nist.gov

**Metal nanoparticle surfaces comprise of multiple planes with various atomic arrangements that interact with gases differently[1,2]. Identification of gas adsorption properties on all facets is an essential prerequisite for rational design of metal nanoparticles for catalysis, energy storage and gas sensing. Adsorbed gas molecules alter the electron density at metal surfaces[3], changing the energy of the surface plasmon resonance[4,5]. All-optical methods using light as the excitation source can identify, *in situ*, gas-metal interactions in an ensemble of metal nanoparticles by measuring energy shifts of either stationary surface plasmon resonances over an entire particle, or delocalized symmetric modes on multiple regions in a particle[6-8]. Such methods preclude the characterization of facet-dependent gas adsorption for individual nanoparticles. Here, by using *in situ* electron-energy-loss spectroscopy in an environmental scanning-transmission electron microscope, we show that localized, stationary surface plasmons on individual facets of triangular crystalline Au nanoparticles in vacuum and in gaseous environments can be excited using a nanometer size electron probe. We then show that, by exploiting this localized spatial resolution, selective gas adsorption on specific sets of facets can be characterized. We anticipate that this method can be extended to quantify the concentration of adsorbed molecules, derive the binding energy for specific facets for certain gas species, and design facet-controlled nanoparticles to achieve specific gas adsorption properties.**

As a fast electron passes by or through a metallic nanoparticle edge, it excites surface plasmons (SP) at the metal surface and loses energy, which can be monitored by electron energy-loss spectroscopy (EELS)[9]. In recent years, electron beam excitation of surface plasmon modes for shape-controlled (cubic and triangular) metal nanoparticles and nanowires has been explored, both experimentally and theoretically [10,11]. Although experiments are consistent with the theoretical predictions, it should be noted that what is actually observed is the excitation, or "feed-in", probability for a given plasmon mode. There is currently no way to map the spatial extent of a surface plasmon directly[11]. For supported nanoparticles, SP modes of different energies are observed at all symmetry-related locations[12,13], suggesting that the energy of a symmetric SP resonance of a supported nanoparticle is delocalized. On the other hand, it has been shown that the local SP energy at the corner of a rectangular nanoparticle surrounded only by vacuum is different from that at the corners in contact with the support[14] i.e. when one corner is exposed to a different dielectric environment from the others. This observation implies that if the dielectric environment, $\varepsilon$, of a part of the particle is different from the rest, then the SP excited by the electron beam may no longer sample the whole particle, but may instead be quite local. It is therefore essential to identify the type of SP resonance (symmetric, whole-particle resonance or local, stationary) excited by an electron beam at specific energies. Importantly, none of the currently available reports define the degree of SP localization and therefore do not determine the spatial resolution of SPs excited by an electron beam. Here, we first estimate the extent to which particular SP excited by an electron beam is localized on a triangular Au particle. Then we exploit the sensitivity of the localized SP energies to small changes in the local dielectric function ($\varepsilon$) to explore gas adsorption on various Au nanoparticle facets.

We measure the local SP energy as a function of the position of an exciting electron beam with respect to a triangular Au nanoparticle supported on $TiO_2$. Measurements are performed using an environmental scanning-transmission electron microscope (ESTEM) equipped with a monochromated 80 keV electron source with an energy resolution of 0.1 eV and a high-resolution electron energy-loss spectrometer with a spectral dispersion of 0.01 eV per pixel. This combination enables the detection of energy loss features above 1 eV and the changes in SP energy with respect to environmental changes, such as supporting substrates and adsorbed gases. Au nanoprisms (≈ 25 nm side-length, ≈ 20 nm thickness) supported on larger TiO$_2$ particles (≈ 100 nm longest side) often exist in a cantilevered configuration, with part of the nanoparticle in close contact with the support and the remainder extending into free space (Fig. 1a). SPs are efficiently excited by positioning the electron-beam proximate to the nanoprism[9,15](Fig. 1a). To explore the spatial extent of electron-induced SPs on our Au nanoprisms, we first probe, with an approximately 1 nm diameter beam, adjacent to the particle edge, sampling along its perimeter sequentially at 2 nm intervals, from a corner that extends into the vacuum (10$^{-5}$ Pa) to one corner that is in contact with the TiO$_2$ support (Figure 1a, 1b). The resulting electron energy-loss spectra (EELS) show a distinct shift in the energy-loss peaks from (2.400± 0.004)[*] eV (Fig. 1c, topmost orange spectrum) to (2.248 ± 0.007) eV (Fig. 1c, bottommost blue spectrum). The (2.400 ± 0.004) eV peak location corresponds to an SPR that minimizes the magnitude of the real and imaginary part of the dielectric function, i.e. the complex function (1 + ε)) for unperturbed Au[16,17], and the peak at (2.248 ± 0.007) eV corresponds to an SP energy modified due to the change in dielectric environment, i.e. TiO$_2$ versus vacuum. The energy shift is not abrupt: broad peaks exist in a transition region that is from 4 nm to 8 nm away from the support, and span the energy range from the unperturbed Au to the support-modified energies (Fig. 1b and 1c, green lines). Peak positions are always at 2.4 eV when probing on the flat side or the curved corner away from the transition zone and the corner in contact with the support. In addition, the support-modified SP energy shift occurs at the one supported corner only while the other corner is unperturbed. These facts suggest that the SP excited here (at 2.4 eV) is not dependent on the geometry of the particle and is a localized, stationary electron oscillation at the surface excited by the electron beam. Based on the size of the region that is affected by the support, we estimate that the lateral spatial extent of the excited SP to be (6 ± 2) nm. The vertical extent may be ≈ 20 nm, assuming the interacting area of electron and facets is approximately equal to the thickness of the nanoprism.

Motivated by this observation of localized electron-induced SPs on our Au nanoprism, we probe, at least 10 nm away from the TiO$_2$ support, the corner location A and side location B. These locations comprise different facets (Fig. 2a). The configuration of the Au nanoparticle on the TiO$_2$ support, as shown in Figure 1a, provides a suitable orientation for electron-beam probing while offering maximal facet exposure to gas molecules and minimal interference from the support. At the base vacuum pressure, $P_0 = 10^{-5}$ Pa, distinct energy-loss peaks are detected at $E_{A,0} = (2.375 \pm 0.004)^*$ eV (Fig. 2b, black symbols) and at $E_{B,0} = (2.389 \pm 0.007)$ eV (Fig. 2c, black symbols) at locations A and B, respectively. SP-EELS collected under 100 Pa H$_2$ ($P_{H_2}$) – a gas of interest for sensing – at location A show a measureable SP energy shift in the negative direction; $\Delta E_{A,H_2} = (-0.071 \pm 0.005)$ eV (Fig. 2b, red symbols). But a negligible shift in SP peak position ($\Delta E_{B,H_2} = (-0.008 \pm 0.006)$ eV) is observed at location B at the same $P_{H_2}$ (Fig. 2c, red symbols). Fully restoring the chamber pressure to its base value $P_0$, in either case, results in the measured SP-EELS peak returning to an average value, $E_{A,0}' = (2.369 \pm 0.008)$ eV (Fig. 2b, green symbols) and $E_{B,0}' = (2.398 \pm 0.006)$ eV (Fig. 2c, green symbols), which are equivalent to initial values of $E_{A,0}$ and $E_{B,0}$ respectively, within the experimental error. The difference in SP energy shifts observed between the corner (A) and side (B) locations upon exposure to H$_2$ is consistent over a range of ≈ 0 Pa to 200 Pa, as shown by measurements of 10 nanoprisms as a function of $P_{H_2}$, at locations A (Fig, 3a) and B (Fig. 3b). The measured shifts, $\Delta E_{A,H_2}$ increase with increasing pressure, reaching a constant value of $\Delta E_{A,H_2} = (-0.07\pm0.013)$ eV† for pressures above $P_{H_2} = 100$ Pa. In contrast, the SP energy shift at location B, $\Delta E_{B,H_2} = (-0.01\pm0.016)$ eV†, is negligible over the same pressure range. The distinctly different $H_2$-induced responses at locations A and B, separated only by ≈ 10 nm, confirm that the gas adsorption-induced shift in SP energy is a local effect and thereby illustrate the high spatial resolution of the SP-EELS method.

---

* The measurement uncertainty for all EELS peaks is the uncertainty of fitting a Gaussian to discrete points of the spectral peak acquired with an energy dispersion of 0.01 eV per pixel.

We further explore the facet dependence of gas absorption, leading to SP energy shifts, using other gases such as CO, which is reported to readily adsorb on the surfaces of Au catalysts used for CO oxidation[18], and $O_2$, which is reported to have only a weak affinity for Au[19,20]. In a CO environment, when probing location A, the SP energy shifts positively as a function of increasing pressure, $P_{CO}$, reaching $\Delta E_{A,CO} = (0.05 \pm 0.014)$ eV†. Over the same $P_{CO}$ range, systematically higher SP energy shifts, reaching $\Delta E_{B,CO} = (0.08 \pm 0.012)$ eV†, are obtained when probing location B. In an $O_2$ environment, no measureable energy shifts are observed as a function of pressure for both locations, which is consistent with the literature. It is interesting to note that, independent of the probe locations, SP energy shifts are found to be of opposite signs for $H_2$ and CO exposure. This is because charge transfer or electrostatic force-induced electron density changes at the metal surface are different depending on the nature of the gas[21]. Hydrogen dissociates on Au surfaces and becomes more negative and the metal becomes more positive. This electron density decrease at the metal surface causes the SP energy to shift to lower values. In contrast, electron transfer occurs from adsorbed CO into the metal *s* and *d* orbitals, leading to an increase in the electron density at the metal surface[22], causing the SP energy to shift positively. The spatial dependence of the observed shift can be attributed to the selectivity of gas adsorption on specific crystal facets[19] present along the edge of the particle. From tomographic STEM images and electron diffraction patterns on a single nanoprism, these surface facets are identified as {100} and {110} at the corner location A and {311} at the side location B of the nanoprism (Fig. 3c, Methods). Assuming a nominal 8 nm SP-EELS lateral resolution, as inferred from Fig. 1, the limited extent (≈ 6 nm) of the {100} and {110} facets at location A, suggests that the spectra recorded there contain, at most, a 25 % admixture of the response from the adjacent {311} facets.

---

† The uncertainty is given by the standard deviation of multiple measurements.

Based on the assumptions of a uniform SP interaction volume and uniform gas adsorption on the sampling surfaces, the SP energy shift is determined by the average electron density change, $\Delta n$, of the probed region. As $\Delta n$ is proportional to the number of adsorbed gas molecules, the measured energy shifts can thus be related to the density of adsorbed molecules and binding energies of the various gas species on individual facets. For example, experimentally measured energy shifts as a function of pressure, $\Delta E[P]$, can be converted into a pressure-dependent local surface electron density change, $\Delta n[P]$ (Supplementary Eqns. 1, 2). The adsorbed molecule density can be obtained using $\Delta n[P]/q$, where $q$ is the charge transfer between a gas molecule and Au surface, and depends on the nature of the gas and the surface atom arrangement of Au. A model based on the isothermal Langmuir equation[23] that describes gas adsorption dynamics for a monolayer at temperature T (Supplementary Eqn. 5), can be fitted to $\Delta n[P]$ for all combinations of gases and facets (Fig. 4). Then the charge transfer density ($q \times \rho$, where $\rho$ is available adsorption site density), and the adsorption-desorption rate constant at equilibrium, $K$, can be derived (Supplementary Table 1). This shows that, as $P_{H_2}$ approaches 0 Pa, the ratio of pressure-dependent adsorption rates for H$_2$ on {100}+{110} facets and {311} facets is 9.91 (Fig. 4a, Supplementary Table 1). This value suggests that H$_2$ is more readily adsorbed on {100}+{110} facets than the {311} facet. As $P_{H_2}$ increases, both $\Delta n_{H_2,\{100\}+\{110\}}$ and $\Delta n_{H_2,\{311\}}$ reach asymptotic values of 5.48 and 0.88, respectively. If we assume that the values of charge transfer for $H_2$ on {100}+{110} and {311} are similar, then the ratio of the given asymptotic values (6.2) will be dependent on the relative density of available surface binding sites. In contrast, CO adsorbs more readily on {311} facets compared to {100} + {110} facets and saturates faster on {311} facets (Fig. 4b, Supplementary Table 1). As $P_{CO}$ increases, the total charge transfer per area reaches asymptotic values of 6.80 and 4.99 for CO on {311} facets and {100} + {110}, respectively. $K$ is controlled by the pre-exponential factor, $A$, and effective binding energy, $E_b$. The effective binding energy for a given gas species on a specific facet can be calculated by knowing that $A$ is related to the gas collision factor[24]. For example, $A$ for CO is assumed to be in the range of $(1 \pm 0.5) \times 10^{-10}$ s$^{-1}$, which yields values of $E_b$ for CO on {311} and {100}+{110} of $(0.52 \pm 0.02)$ eV and $(0.49 \pm 0.02)$ eV, respectively, which are comparable to the values in theoretical calculations[25].

In summary, we have shown that the stationary SPs induced by the highly localized electric field of an electron beam sample the dielectric properties of a relatively small area proximate to the beam. By taking advantage of SP sensitivity to small changes in the electron density, we show that SP-EELS can be used to detect gas adsorption on selected facets of a nanoparticle. Quantitative measurements are possible by correlating experimental SP energy shifts to electron density changes and knowing the charge transfer for gas adsorption on the surfaces. The observed selectivity of gas adsorption for certain surfaces is due to the difference in surface atomic arrangement. For example, significant $H_2$-induced energy shifts at the Au corner indicates that hydrogen can interact strongly with {100} and {110} facets. It is interesting to note that $H_2$ adsorption on Au cannot be directly measured when using all-optical methods due to the ambiguity of the collective response from an ensemble of particles as well as from the changes in the dielectric constant of the substrate (e.g. support reduction)[26]. By contrast, we have presented a technique with high spatial resolution that allows examination of selected facets independently. We anticipate exploiting the potential of SP-EELS to classify the facets that promote selective gas adsorptions, thereby enabling the design of nanoparticles with facet distributions that enhance their performance in various applications.

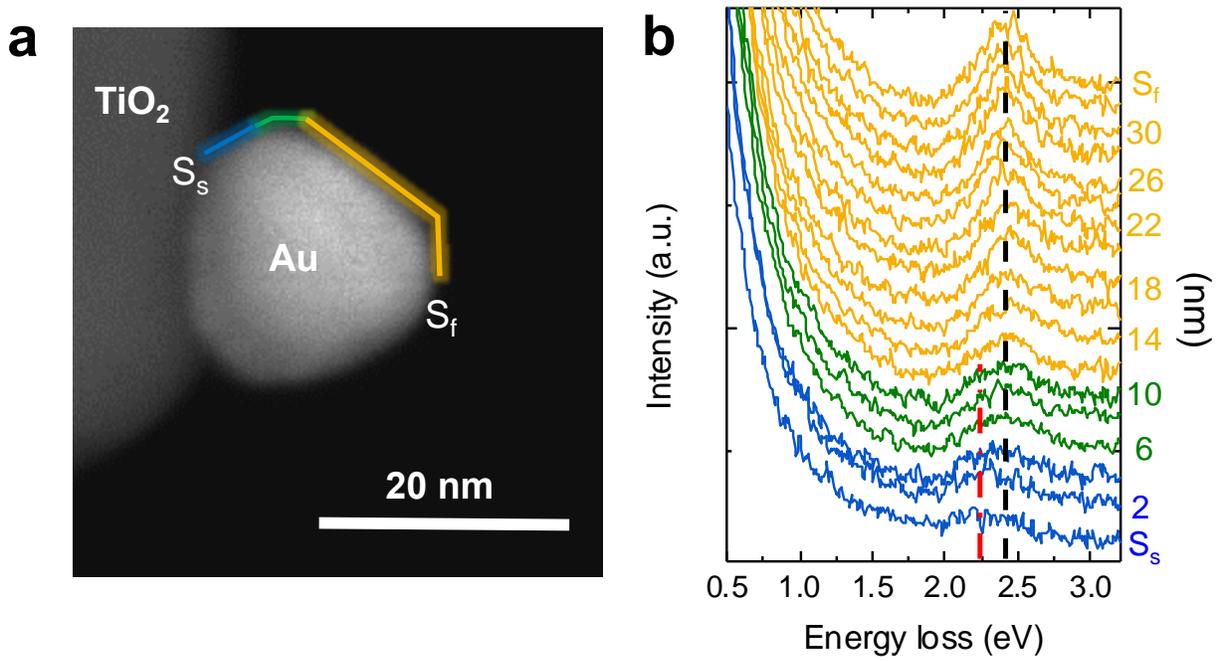

**Figure 1| Local SP energy shift reveals by electron beam excitation nearby the TiO₂-modified edge of Au nanoprism. a,** A STEM image of a Au nanoprism (side length 25 nm), with a plate surface normal to the beam axis, attached by one edge to a TiO$_2$ support. Electron beam (≈ 1 nm diameter) sequentially probes locations along the Au perimeter from one corner in vacuum to another corner that is in contact with TiO$_2$ support. The path starts from S$_s$, along blue, green and yellow lines, and ends at S$_f$. **b,** Corresponding EEL spectra for each location in Figure 1a that show peaks at 2.400 eV for the yellow region, 2.248 eV for the blue region and peak-shifts in between for the green region. All spectra intensities are normalized to the zero-loss peak intensity and vertically displaced for clarity. Black and red dotted lines are positioned at 2.400 eV and 2.248 eV, respectively.

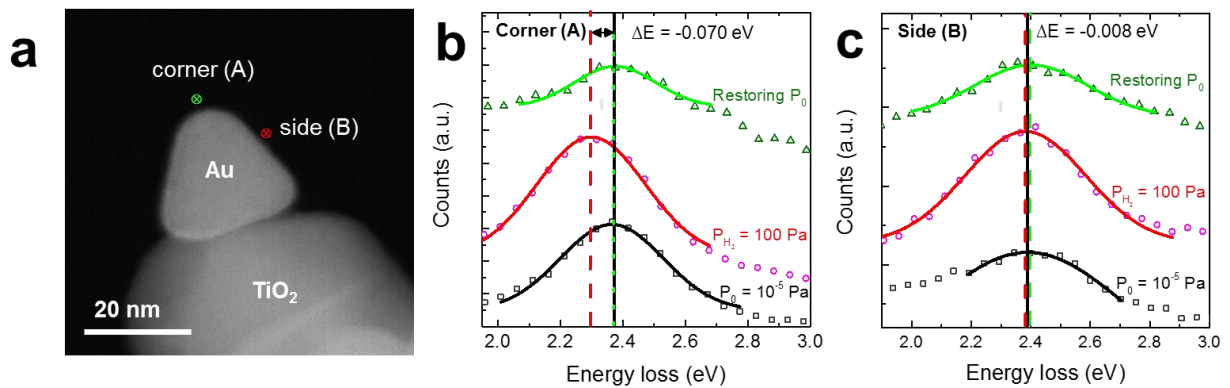

**Figure 2| Electron beam induced SP as a nano-scale probe to detect facet-dependent gas-metal interaction on an individual Au nanoprism. a,** A STEM image of Au nanoprism with a plate surface normal to the beam axis, attached to one edge of a TiO₂ support. Location A is the midpoint of the corner {110} + {100} facets, and location B is the midpoint of the side {311} facets. **b,** A stack of EEL spectra collected at the corner location A and (**c**) the side location B under $P_0$ (black), $P_{H_2}$ = 100 Pa (red), and restored $P_0$ (green). All spectra intensities were normalized to the zero loss peak intensity and vertically displaced for clarity. A Gaussian function is fit to individual spectra and the peak location represents the SP energy. Black, red and green dotted lines are positioned at the SP energies measured at $P_0$, $P_{H_2}$ = 100 Pa and restored $P_0$, respectively.

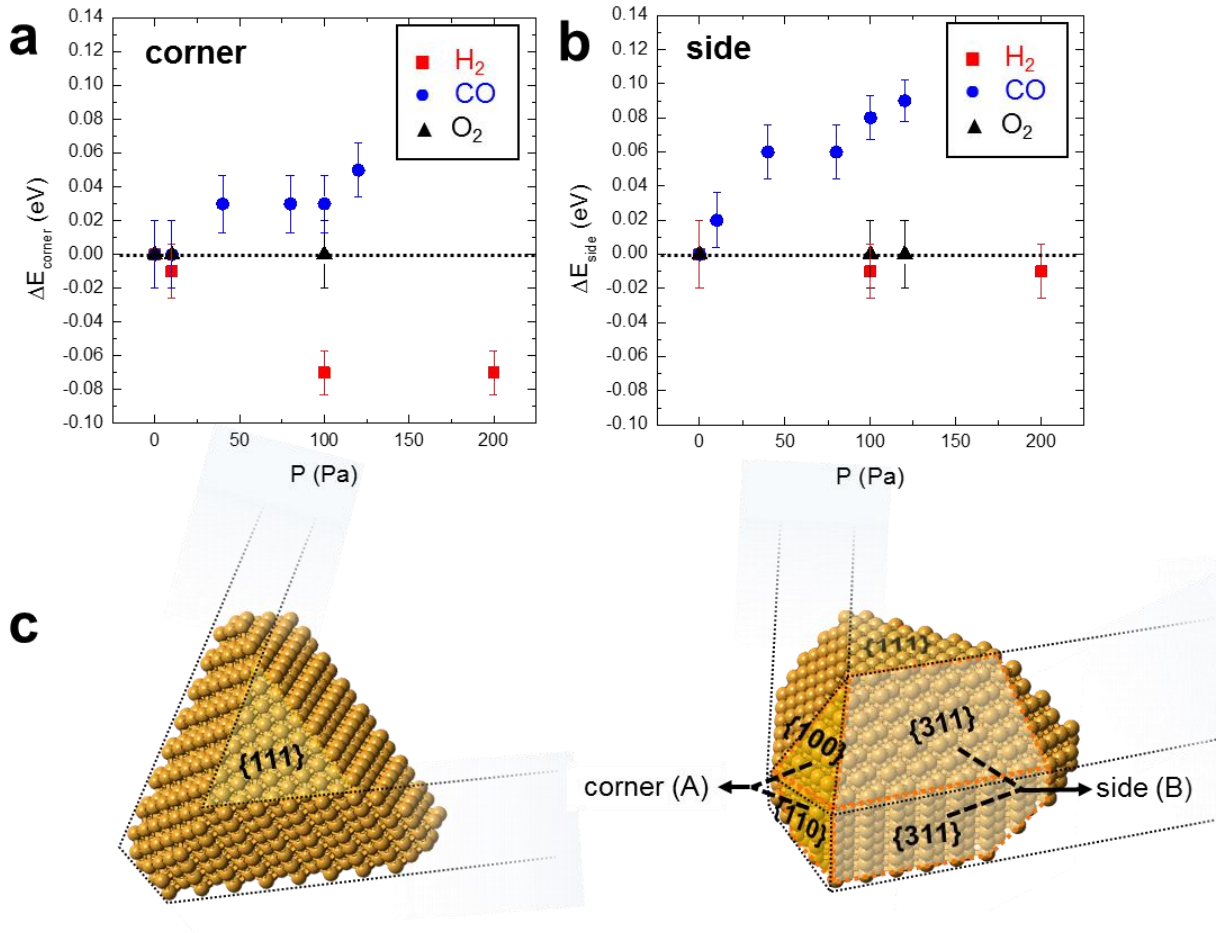

**Figure 3| Different signs and extents of LSPR energy shifts (ΔE) as a function of gas pressure $P$, for six combinations of gas type and Au facet (corner location A : {100}+{110} facets and side location B : {311} facets).** Measured Δ$E$ as a function of gas pressure of H₂, CO, and O₂ (ranging from 10⁻⁵ Pa to 200 Pa) when probing at (**a**) corner location A and (**b**) side location B. Error bars indicate single standard deviation uncertainties of multiple measurements. **c,** Idealized atomic models of nanoprism: viewing direction along the beam going into the paper (left),

triangular plate surface of Au nanoprism is {111}; viewing direction perpendicular to the beam trajectory (right), corner location A is comprises {100} and {110} facets and side location B comprises {311} facets.

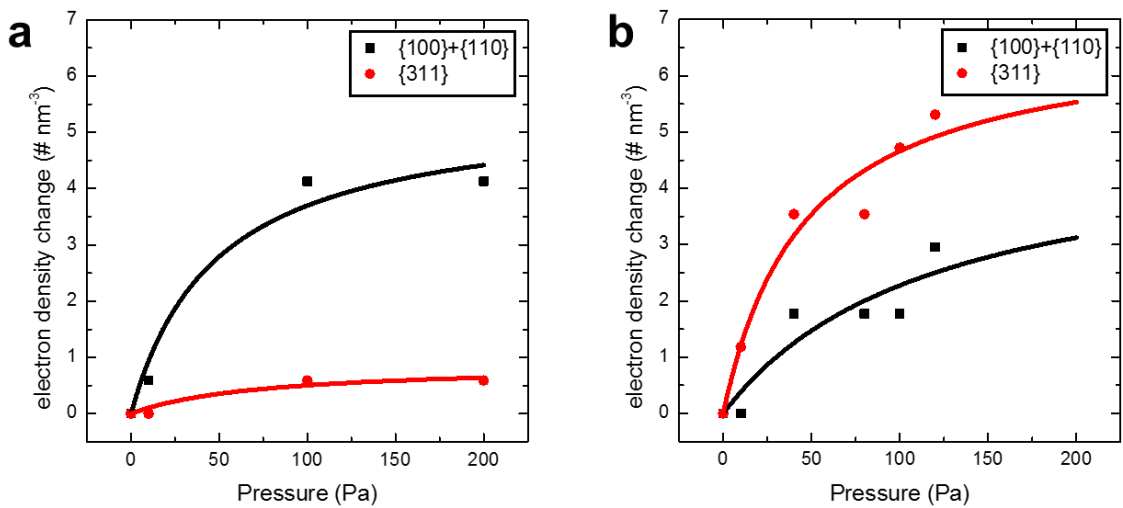

**Figure 4| Preferential gas adsorption on selective surfaces of an individual Au nanoprism.** Estimated electron density change (from measured energy SP shift) indicates gas adsorption quantity as a function of gas pressure for the corner ({100}+{110}) and the side ({311}) facets. Fits to a Langmuir absorption isotherm (solid lines) are plotted for $H_2$ (**a**) and CO (**b**).

## Methods

### Materials

Au nanoprisms of side length ≈ 25 nm and thickness ≈ 20 nm, were synthesized in solution by seed-mediated method[27]. Small Au seeds (< 2 nm) were first prepared by reducing an Au precursor, $HAuCl_4 \cdot 3H_2O$ ($2.5 \times 10^{-4}$ mol·L$^{-1}$), with ice-cold $NaBH_4$ ($1.8 \times 10^{-3}$ mol·L$^{-1}$), using cetyltrimethylammonium bromide (CTAB) ($22.5 \times 10^{-2}$ mol·L$^{-1}$) as a surfactant. This seed solution was magnetically stirred for 2 min, and set aside for a duration of at least 2 hr (but no greater than 12 hr). A fraction of the Au seed solution (0.01 ml) was then enriched by adding 0.2 ml of $HAuCl_4 \cdot 3H_2O$ ($2.0 \times 10^{-4}$ mol·L$^{-1}$), along with 1.6 ml of CTAB ($1.6 \times 10^{-2}$ mol·L$^{-1}$), and 0.6 ml of ascorbic acid ($6.0 \times 10^{-3}$ mol·L$^{-1}$). Commercial rutile $TiO_2$ nanoparticles (≈ 100 nm longest side) were added to the resulting Au nanoparticle solution after it was washed using distilled water to remove left-over Au precursors, CTAB and ascorbic acid. The solution that contains Au nanoparticles and $TiO_2$ nanoparticles (3 µl) was drop-cast on a lacy carbon-coated copper TEM grid. As-grown particles consisted mostly of Au nanoprisms, based on imaging with a TEM[28]. Nanoprisms were plasma cleaned to remove surfactants from their surfaces before all experiments. This is a critical step since SPR energy is highly sensitive to the surrounding medium, and a clean surface of Au for gas adsorption and chemical reaction is required.

A miniature 3D reconstructed crystal model in 1:5 scale was produced using a combination of tomography images, high resolution TEM (HRTEM) images and single Au nanoprism electron diffraction pattern, (Supplementary Fig. 1). This model provides crystallographic information for all surfaces.

**Surface plasmon measurements**

Electron energy loss spectra (EELS) were collected in a spectrometer with a spectral dispersion of 0.01 eV per pixel, in scanning transmission electron microscope (STEM) mode at 80 kV with an approximately 1 nm diameter beam, using an environmental scanning transmission electron microscope (ESTEM). The microscope is equipped with a monochromator, which improves the energy resolution to 0.1 eV (defined at full-width half maximum, with 0.2 eV at full-width tenth maximum) and a differentially pumped sample chamber attached to a multiple gas injection system (MGIS), which enables exposure of the sample to various gasses up to a maximum local pressure of $\approx$ 2000 Pa[29].

Measurements were performed on 10 Au nanoprisms of similar size each subjected to one cycle of gas injection with increasing pressures and vacuum ($P_0$) restoration. Each spectrum was obtained by summing 100 acquisitions of 10 ms exposures, after aligning to the zero loss peak (ZLP), to enhance signal to noise ratio without saturating the charge-coupled device (CCD) camera. In order to find the accurate energy loss peak position, the ZLP was removed from the energy-loss spectra by fitting to a Gaussian and squared Lorentzian function, and then the residual peak in between 2 to 3 eV was fitted with a Gaussian function. The center of the Gaussian peak was defined as the SP energy.

# Probing gas adsorption on individual facets of a metal nanoparticle

Pin Ann Lin[1,2], Jonathan Winterstein[1], John Kohoutek[1,2], Henri Lezec[1] and Renu Sharma[1*]

Affiliation:

[1]Center for Nanoscale Science and Technology, National Institute of Standards and Technology, Gaithersburg, MD 20899-6203, USA

[2]Maryland NanoCenter, University of Maryland, College Park, MD 20742, USA

*Corresponding author email:

Renu.sharma@nist.gov

## Supplementary Information

**Supplementary Methods**

The free electron concentration in a region $N$ (e.g. corner or side) is independently specified according to:

$$n_N = n_{Au} + \Delta n_N \qquad (1)$$

The local plasma frequency is given by:

$$\omega_{p,N} = \sqrt{\frac{n_N e^2}{\varepsilon_0 m^*}} \qquad (2)$$

Using the effective free electron mass in Au, ($m^*$), and the dielectric function of Au ($\varepsilon_0$). The value of $\omega_{p,N}$ is proportional to square-root of $n_N$, thus $\omega_{p,N}$ increases (decreases) when $n_N$ increases

(decreases). The change in electron density ($\Delta n_N$) can therefore be calculated from the observed SP energy shift ($\Delta \omega_{p,N}$).

**Gas adsorption model: Langmuir isotherm function**

We model surface gas adsorption on a given Au nanoparticle facet by using a Langmuir isotherm function[23] given by:

$$\theta(P) = \frac{KP}{1+KP} \qquad (3)$$

where $\theta$ is the fraction of occupied surface sites, $P$ is gas pressure and $K$ is the ratio of adsorption and desorption rate constants at equilibrium. The molecular surface coverage per area is given by:

$$N(P) = \theta(P)\rho \qquad (4)$$

where $\rho$ is the available adsorption site density. Gas adsorption leading to such surface coverage also yields a change in surface electron density given by:

$$\Delta n(P) = qN(P) = q\rho \frac{KP}{1+KP} \qquad (5)$$

A least squares fit of eq. 5 to estimates $\Delta n$ as a function of pressure $P$, derived from the experimental data, yields the following parameters: $q$ (charge transfer), $\rho$, and $K$, for each combination of facet and gas type. $K$ and $q \times \rho$ are given in Table S1., along with the quantity $q \times \rho \times K$ representing the slope of $\Delta n(P)$ at the origin.

Here $K$ depends on an effective surface binding energy $E_b$ (which is facet-dependent) according to:

$$K = A\exp(-E_b/kT) \tag{6}$$

where $A$ is a pre-exponential factor that depends on the nature of the gas interaction with specific surface atomic arrangement, $k$ is Boltzmann's constant, and $T$ is temperature.

**Supplementary Table 1**. Product of charge transfer and available adsorption site density, $q \times \rho$, adsorption-desorption rate constant at equilibrium, $K$, and pressure dependent adsorption rate, $q \times \rho \times K$ for four combinations of gas types and Au facets obtained by fitting a Langmuir function to experimentally estimated $\Delta n[P]$.

| Gas type on Au facets | $q \times \rho$ (charge per nm$^2$) | $K$ | $q \times \rho \times K$ (charge per nm$^2$ per Pa) |
|---|---|---|---|
| H$_{2\{100\}+\{110\}}$ | 5.48 | 0.0207 | 0.113 |
| H$_{2\{311\}}$ | 0.88 | 0.013 | 0.011 |
| CO$_{\{100\}+\{110\}}$ | 4.99 | 0.008 | 0.039 |
| CO$_{\{311\}}$ | 6.80 | 0.0217 | 0.147 |

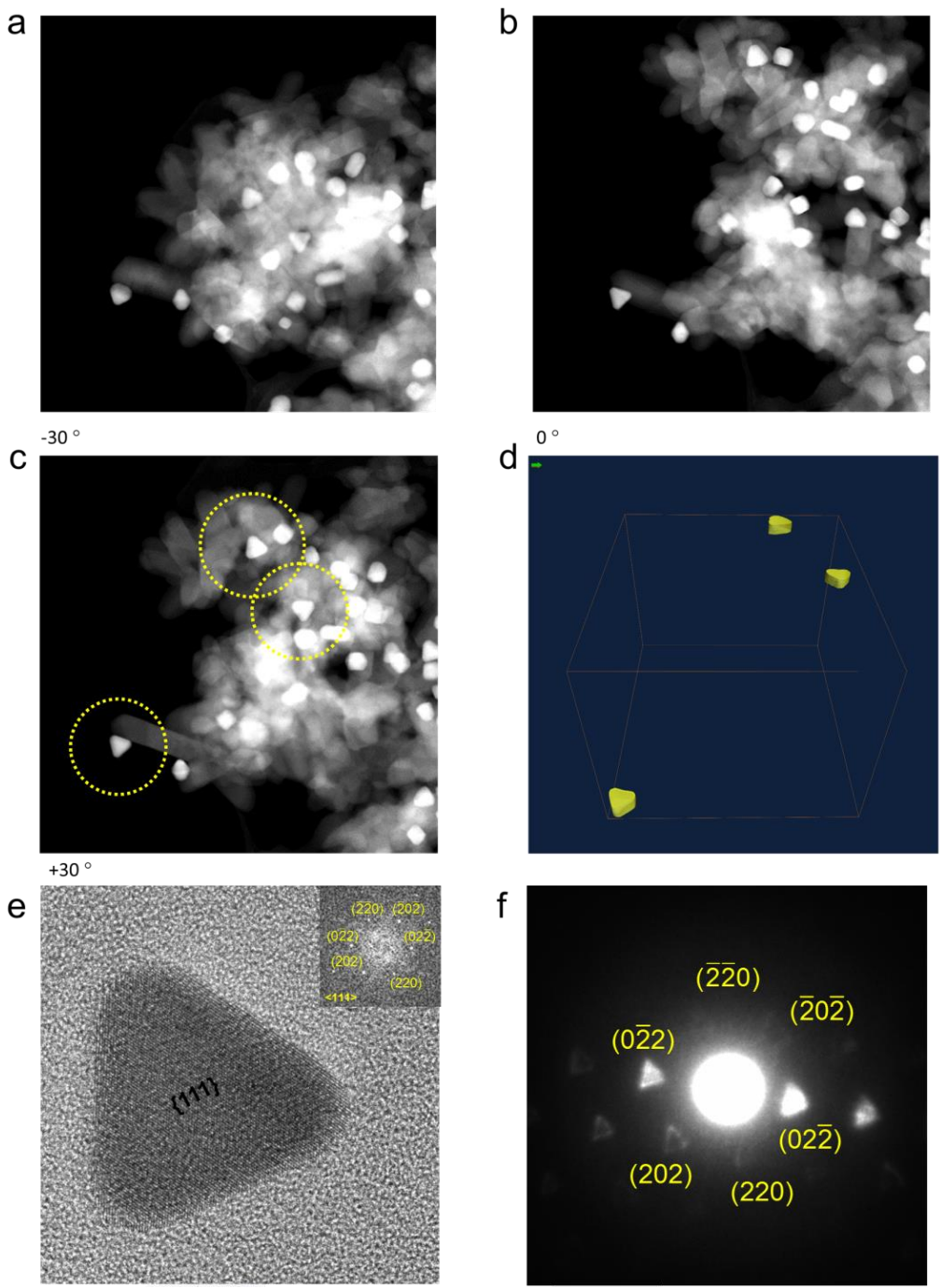

**Supplementary Figure 1**. STEM tilt series images of Au on TiO$_2$ at (**a**) -30°, (**b**) 0 °, and (**c**) 30 ° for (**d**) 3D reconstruction. **e,** HRTEM image of Au nanoprism with a diffractogram to show the Au$_{FCC}$ along the <111> zone axis (inset). The plate surface is identified as {111}. **f,** Nanobeam diffraction pattern, showing multiple dark field images, of the nanoprism along the <111> zone axis.